\documentclass[%
 reprint,
 amsmath,amssymb,
 aps,
 prl,
 longbibliography,
 lengthcheck,%
]{revtex4-1}

\usepackage{graphicx}
\usepackage{dcolumn}
\usepackage{bm}
\usepackage{hyperref}

\begin{document}

\preprint{APS/123-QED}

\title{Waves of magnetic moment and generation of waves by neutron
beam in quantum magnetized plasma
}

\author{P. A. Andreev}%
\email{andreevpa@physics.msu.ru}
\affiliation{
 Department of General Physics,
 Faculty of Physics, Moscow State
University, Moscow, Russian Federation.}
\author{L. S. Kuzmenkov}%
 \email{lsk@phys.msu.ru}
\affiliation{%
 Department of Theoretical Physics, Physics Faculty, Moscow State
University, Moscow, Russian Federation.}%




\date{\today}

\begin{abstract}
This paper is devoted to studying of dispersion of waves in the
magnetized plasma with the spin and exploring of new methods of the
generation wave in the plasma. We consider the dispersion of waves,
existed in the plasma in consequence of dynamic of the magnetic moments.
It is shown there are nine new waves in the magnetized plasma because
of the magnetic moments dynamic. We show there are instabilities at
propagation of the neutron beam through the plasma. Increments of
instabilities caused by neutron beam are calculated. For studying
of this effects we generalize and use the method of the many-particle
quantum hydrodynamics. Described processes can play important role at
calculation of the stability and the safeness of the nuclear reactors and the studying of the processes in the
atmosphere of the neutron stars.
\end{abstract}

\pacs{...}
\maketitle


\section{I. Introduction}

In plasma physics the effect of generation of waves by means of an
electron beam is well known ~\cite{Miller
BOOK}-~\cite{Mikhailovskii BOOK} and some aspects of this problem have been studied ~\cite{Bret
PRL 08}-~\cite{Bret PP 08}. The
self-consistent electromagnetic interaction of the electrons of
the beam with the electrons and ions of the plasma plays a crucial role. The
increasing of an amplitude of collective oscillation arises. At these processes  the generation of waves
takes place in both cases at the propagation of the beam through the
plasma and at the propagation of the beam along a space-limited medium due to the
long-range electromagnetic interaction.

Plasma is the isotropic medium. At presence of the uniform
external magnetic field the preferential direction is arisen -- the
direction of the external magnetic field. In this case the properties
of the plasma change essentially. Dispersion dependence of waves changes. Besides new types of waves emerge. The dispersion properties
of these waves essentially depend on the angle between the direction of the wave
propagation and the direction of the external magnetic field.

In our work we consider the plasma in an external uniform magnetic
field. We regard the motion of both the electrons and the
ions. We pay attention to the spin
of particles, more precisely their magnetic moments. The dynamic of
magnetic moments leads to change in the dispersion dependence of
excitations, which exist in the plasma at the absence of magnetic moments. It also leads to arising of new waves. The
two waves arise at propagation along the external magnetic field
and the four waves emerge at propagation perpendicular to the external field.

Special attention, we give to the process of the interaction of
a neutron beam with the plasma. Neutrons have magnetic moment.
Consequently, they can interact with the spins and the currents in the plasma by
means of the collective magnetic field. In that process, the beam of
neutrons transfers energy to the plasma and generates waves. We focus our attention on case when the neutron beams
moving in the direction of the external magnetic field. We also consider
an opportunity of generation of the waves which propagate along the
external magnetic field.

In last years there have been a lot of papers devoted to influence of spin
on dynamic of the plasma. Brief review of this studies is presented in
Ref.s ~\cite{P.K. Shukla UFN 10,Shukla RMP 11,Andreev arxiv 11 3}. Recently, the
quantum kinetic studying of the waves in the plasma was made ~\cite{Asenjo arxiv 2011}.
Contribution of the semi-relativistic effects was studied there. Another examples of
derivation of the quantum kinetic equation are considered in Ref.s
~\cite{Maksimov TMP 2002,Tyshetskiy 2011}.

Moreover, we make further development of the method of investigation of similar problem, which is called the many-particle quantum hydrodynamics.

The Vlasov kinetic equation or the corresponding hydrodynamic
equations for system of charged particles is used in the classic
physics at investigation of plasma properties.

Consistent investigation of dynamics of magnetic moment or spin of
particles expects of the quantum-mechanical description. Consequently,
we derive equations of the collective quantum motion. We obtain them from the many-particle Schrodinger equation.
They arise in the form similar to the hydrodynamics equations.
Therefore, the developing method is called quantum hydrodynamics (QHD). For the first
time, for many-body system, this method was proposed in paper
~\cite{MaksimovTMP 1999}. Its generalization for the system of
spinning particles was made in Ref. ~\cite{MaksimovTMP 2001}.
Another method of derivation of the QHD equations is proposed in paper
~\cite{Marklund PRL07}. Relativistic generalization of this
approximation was made in Ref. ~\cite{Asenjo PP 11}.
We study the contribution of the spin-current and the spin-orbit
interactions. The spin-current interaction is the interaction between the
moving charge and the motionless spin, more precisely magnetic
moments. The spin-orbit interactions is the interaction between the
motionless charged particles and the moving spin. For the system of
two particles one of the described interactions could be exclude. We
can make it by using appropriate frame. But for many-body system
we need to consider both of them. The system of the quantum
hydrodynamic equations in the self-consistent field approximation looks like
the same obtained from the Schrodinger equation for one particles in
external field. The many-body QHD contains a quantum
correlation. We do not consider correlation in this paper. Good
example of calculation of the correlations is presented in Ref.
~\cite{Andreev PRA08}, where evolution of neutral quantum
particles is studied. In this case there is no self-consistent
field and interaction determined by the correlations only.

We make further development of fundamental results obtained in
Ref.s ~\cite{MaksimovTMP 2001} and ~\cite{Andreev AtPhys 08}. We
generalize method of the QHD ~\cite{MaksimovTMP 2001} including the
spin-current and the spin-orbit interactions along with the spin-spin
and the Coulomb interactions between particles. Effect of
 increasing of instabilities at the propagation of the neutron beam
through the magnetized plasma was predicted in ~\cite{Andreev AtPhys
08}. Magnetic moment of the neutron plays crucial role in this effect.
This effect is consequence of the spin-spin and the spin-current
interaction between the spins of the beam \textit{and} the spin and current of the
plasma. Here we study the influence of the spin-orbit interaction on the
described effect. This effect can be called the effect of the resonant
interaction of the neutron beam with the plasma. In the part of the paper,
where we consider the propagation of the beam through the medium, we bound
oneself by the case there both the beam and waves propagate along the
external magnetic field.

We are interested in instabilities which arise at interaction of the beam with
both the well-known magneto-plasma and the recently predicted waves
~\cite{Andreev VestnMSU 2007}, ~\cite{Andreev PIERS 2011}. In this work we
do not present the amendments, caused by the spin, in the dispersion of the well-known
waves. Readers who are interested in this subject see Ref.s ~\cite{P.K. Shukla UFN 10}
and ~\cite{Andreev VestnMSU 2007}.

This paper is organized as follows.  In Sec. II we present the
system of the quantum hydrodynamic equations.   In Sec. III we
describe the method of solving of the equations for the case of linear
excitations. In Sec. IV we study dispersion of new type of waves in the
plasma caused by dynamic of magnetic moments. We examine the waves
propagated along the external magnetic field. We research propagation
of the neutron beam through the plasma. We consider interaction of the
beam with both the well-known plasma waves and the new waves caused by spin
dynamics. We show existence of
instabilities in this case and calculate the increments of the
instabilities. In Sec. V we consider waves propagated perpendicular
to the external magnetic field. We interested in dispersion of waves
caused by spin dynamics. In Sec. VI the dispersion of the spin
waves is obtained. We make special attention for  the spin waves dispersion
in the whole $\textbf{k}$ space. In Sec. VII we present the brief summary of our
results.

\section{II. The model}

The equations of quantum hydrodynamic are derived from the
non-stationary Schrodinger equation for system of N particles:
$$\imath\hbar\partial_{t}\psi(R,t)=\Biggl(\sum_{n}\biggl(\frac{1}{2m_{n}}\textbf{D}_{n}^{2}+e_{n}\varphi_{n,ext}\biggr)$$
\begin{equation}\label{NBS Hamiltonian}+\sum_{n,k\neq
k}\frac{1}{2}e_{n}e_{k}G_{nk}\Biggr)\psi(R,t).\end{equation} The
following designations are used in the equation (\ref{NBS
Hamiltonian}):
$D_{n}^{\alpha}=-\imath\hbar\partial_{n}^{\alpha}-e_{n}A_{n,ext}^{\alpha}/c$,
$\varphi_{n,ext}$, $A_{n,ext}^{\alpha}$ - the potentials of the
external electromagnetic field,
$$\partial_{n}^{\alpha}=\nabla_{n}^{\alpha}=\frac{\partial}{\partial x_{n,\alpha}},$$
and $G_{nk}=1/r_{nk}$, - is the Green functions of the Coulomb
interaction, $\psi(R,t)$-is psi function of N particle system,
$R=(\textbf{r}_{1},...,\textbf{r}_{N})$, $e_{n}$, $m_{n}$-are the charge and the mass of particle, $\hbar$-is the Planck constant and $c$ is the speed of light.

We consider the interaction between particles and action
of the external electromagnetic field on them.

The method of the QHD is based on the Schrodinger
equation, which is the non-relativistic. But the Schrodinger equation
allows to account the semi-relativistic amendments, accurate to
$v^{2}/c^{2}$. In this work we interested in the spin-spin, the
spin-current and the spin-orbit interactions along with the Coulomb
interaction. In this section we present the Hamiltonian included the
Coulomb interaction only. We made it for simplicity. Resulted
equations, see below in this section, contain all described
interactions. Whole Hamiltonian is presented in the Appendix.

Here we briefly present the basic steps of derivation of the QHD equation. More detailed description of derivation of the
many-particle QHD equations can be find in ~\cite{MaksimovTMP 2001}.

The first step in derivation of equations of the QHD is a definition of concentration of particles.
We determine the concentration of particles as quantum-mechanical
average of the operator of concentration:
$$\hat{n}=\sum_{n}\delta(\textbf{r}-\textbf{r}_{n}).$$
This function is the microscopic function of concentration in
classic physics, $\delta(\textbf{r})$-is the Dirac's
$\delta$-function.

In that way, the concentration has form:
\begin{equation}\label{NBS def density}n(\textbf{r},t)=\int dR\sum_{n}\delta(\textbf{r}-\textbf{r}_{n})\psi^{*}(R,t)\psi(R,t),\end{equation}
where $dR=\prod_{n=1}^{N}d\textbf{r}_{n}$.

Differentiating the concentration with respect to time, using the
Schrodinger equation and evident form of the Hamiltonian, we obtain the continuity equation:
\begin{equation}\label{NBS continuity equation}\partial_{t}n(\textbf{r},t)+\nabla\textbf{j}(\textbf{r},t)=0.\end{equation}
In that equation a function of current  $\textbf{j}(\textbf{r},t)=n(\textbf{r},t)\textbf{v}(\textbf{r},t)$  is arisen, where
  $\textbf{v}(\textbf{r},t)$- is the velocity field.

Differentiating the function of current with respect to time, we obtain the momentum balance equations, this equation is an analog of the Euler equation:
$$mn(\textbf{r},t)(\partial_{t}+v^{\beta}(\textbf{r},t)\nabla^{\beta})v^{\alpha}(\textbf{r},t)+\partial_{\beta}p^{\alpha\beta}(\textbf{r},t)$$
$$-\frac{\hbar^{2}}{4m}\partial^{\alpha}\triangle
n(\textbf{r},t)+\frac{\hbar^{2}}{4m}\partial^{\beta}\Biggl(\frac{\partial^{\alpha}n(\textbf{r},t)\cdot\partial^{\beta}n(\textbf{r},t)}{n(\textbf{r},t)}\Biggr)$$
$$=en(\textbf{r},t)E^{\alpha}(\textbf{r},t)+\frac{e}{c}\varepsilon^{\alpha\beta\gamma}n(\textbf{r},t)v^{\beta}(\textbf{r},t)B^{\gamma}(\textbf{r},t)$$
\begin{equation}\label{NBS momentum balance eq}+M^{\beta}(\textbf{r},t)\nabla^{\alpha}B^{\beta}(\textbf{r},t)+F_{s-o}^{\alpha}(\textbf{r},t),
\end{equation}
where $\textbf{E}$ and $\textbf{B}$ are the electric and magnetic fields, $\textbf{M}$ is the density of magnetic moments, $\varepsilon^{\alpha\beta\gamma}$ - is the antisymmetric symbol (the Levi-Civita symbol), $p^{\alpha\beta}$ is the kinetic pressure tensor.

Terms which proportional to $\hbar^{2}$  is the quantum Bohm potential, they appear as a result of using of quantum mechanics. In right-hand side of equation (\ref{NBS momentum balance eq}) a force field locates. The force field consists of the Lorenz force, force of interaction of spins with magnetic field and spin-orbit interaction. Form of force which act on the magnetic moments is differ from analogous in the mechanics, which has the form of $\textbf{F}_{mec}=(\textbf{M}\nabla)\textbf{B}$.

The force field for the spin-orbit interaction $\textbf{F}_{s-o}$ arises in the form:
$$F_{s-o}^{\alpha}(\textbf{r},t)=\frac{1}{c}\frac{2\gamma}{\hbar}\varepsilon^{\alpha\beta\mu}\varepsilon^{\beta\gamma\delta}B^{\gamma}(\textbf{r},t)M^{\delta}(\textbf{r},t)E^{\mu}(\textbf{r},t)$$
$$-\frac{1}{c}\varepsilon^{\alpha\beta\gamma}M^{\beta}(\textbf{r},t)\partial_{t}E^{\gamma}(\textbf{r},t)$$
$$-\frac{1}{c}\varepsilon^{\alpha\beta\gamma}\partial^{\delta}E^{\gamma}(\textbf{r},t)J_{M}^{\beta\delta}(\textbf{r},t)$$
\begin{equation}\label{NBS s-o int force}-\frac{1}{c}\varepsilon^{\beta\gamma\mu}J_{M}^{\beta\gamma}(\textbf{r},t)\partial^{\alpha}E^{\mu}(\textbf{r},t),\end{equation}
where $J^{\alpha\beta}_{M}(\textbf{r},t)$ - is the tensor of current of magnetic moment.

In this paper we obtain contribution of the spin-orbit interaction in the Euler equation and equation of balance of magnetic moment.

The equation of evolution of the magnetic moment is:
$$\partial_{t}M^{\alpha}(\textbf{r},t)+\nabla^{\beta}J^{\alpha\beta}_{M}(\textbf{r},t)$$
$$=\frac{2\gamma}{\hbar}\varepsilon^{\alpha\beta\gamma}\Biggl(M^{\beta}(\textbf{r},t)B^{\beta}(\textbf{r},t)$$
\begin{equation}\label{NBS magn mom balance eq}+\frac{1}{c}\varepsilon^{\beta\mu\nu}J^{\gamma\nu}_{M}(\textbf{r},t)E^{\mu}(\textbf{r},t)\Biggr).\end{equation}
This equation is analog of the Bloch equation. Last term is the contribution of the spin-orbit interaction.

This equations take place for each species of particles. They connected by means of the Maxwell's equations:
$$\begin{array}{ccc} \nabla\textbf{B}(\textbf{r},t)=0 ,& \nabla\textbf{E}(\textbf{r},t)=4\pi\sum_{a}e_{a}n_{a}(\textbf{r},t) \end{array},$$
$$ \nabla\times\textbf{E}(\textbf{r},t)=-\frac{1}{c}\partial_{t}\textbf{B}(\textbf{r},t) $$
$$\nabla\times\textbf{B}(\textbf{r},t)=\frac{1}{c}\partial_{t}\textbf{E}(\textbf{r},t)$$
\begin{equation}\label{NBS Maxwell eq}+\frac{4\pi}{c}\sum_{a}e_{a}n_{a}(\textbf{r},t)\textbf{v}_{a}(\textbf{r},t)+4\pi \sum_{a}\nabla\times\textbf{M}_{a}(\textbf{r},t),\end{equation}
where subindex "a" describe the kind of particle: electrons, ions and neutrons.

The tensor of the magnetic moments flow $J_{M}^{\alpha\beta}$ may
be approximately presented in the form
$J_{M}^{\alpha\beta}=M^{\alpha}v^{\beta}$. This is true if we do not
consider the thermal motion of magnetic moments.

\section{ III. Dispersion equation}

We consider the small perturbation of equilibrium state like
$$\begin{array}{ccc}n_{a}=n_{0a}+\delta n_{a}, &\textbf{E}=0+\textbf{E} \end{array},$$
$$\begin{array}{ccc}\textbf{B}=B_{0}\textbf{e}_{z}+\delta \textbf{B},&
\textbf{v}_{a}=0+\textbf{v}_{a} \end{array},$$
$$\begin{array}{ccc}\mu_{a}^{\alpha}=\mu_{0a}^{\alpha}+\delta \mu_{a}^{\alpha},& M_{0a}^{\alpha}=n_{0a}\mu_{0a}^{\alpha}=\chi_{a}
B_{0}^{\alpha},\end{array}$$
\begin{equation}\label{NBS linearisation}\begin{array}{ccc}
n_{0e}=n_{0i},& p^{\alpha\beta}_{a}=p_{a}\delta^{\alpha\beta}, &
\delta p_{a}=m_{a}v_{sa}^{2}\delta n_{a},
\end{array}\end{equation}
here $\delta^{\alpha\beta}$- is the Kronecker symbol,
$\chi_{a}=\kappa_{a}/\nu_{a}$ is the ratio between the equilibrium
magnetic susceptibility $\kappa_{a}$ and magnetic permeability
$\nu_{a}=1+4\pi\kappa_{a}$. In the case there $\kappa_{a}\ll 1$ we
have $\chi_{a}\simeq\kappa_{a}$. In equations (\ref{NBS
linearisation}), $v_{sa}^{2}$ is the thermal velocity, for the case
of degenerate electrons $v_{sa}^{2}$ is the Fermi velocity.
Substituting these relations into the system of equations (\ref{NBS
continuity equation}), (\ref{NBS momentum balance eq}), (\ref{NBS
magn mom balance eq}) and (\ref{NBS Maxwell eq}) and neglecting by
nonlinear terms, we obtain a system of linear homogeneous
equations in partial derivatives with constant coefficients.
Passing to the following representation for small perturbations
$\delta f$
$$\delta f =f(\omega, \textbf{k}) exp(-\imath\omega+\imath \textbf{k}\textbf{r}) $$
 yields a homogeneous system of algebraic equations.
The electric field strength is assumed to have a nonzero value.
Expressing all the quantities entering the system of equations in
terms of the electric field, we come to the equation
\begin{equation}\label{NBS general form of disp eq} \Lambda_{\alpha\beta}(\omega,\textbf{k})E_{\beta}(\omega,\textbf{k})=0, \end{equation}
Since the amplitude of the electric field is not equal to zero we have
the dispersion equation
$$Det\widehat{\Lambda}(\omega,\textbf{k})=0.$$
We do not present the explicit form of $\Lambda_{\alpha\beta}(\omega,\textbf{k})$
because of its largeness.

\section{ IV. The waves propagated parallel to magnetic field}

In the absence of the neutron beam we have the magnetized plasma of particles
with magnetic moment. Dispersion equation have two new solutions:
\begin{equation}
\omega=\mid\Omega_{e}\mid\Biggl(1+\frac{8\pi
k_{z}^{2}c^{2}\chi_{e}}{\omega_{Le}^{2}+2k_{z}^{2}c^{2}-2\Omega_{e}^{2}}\Biggr),
\label{NBS new el par branch}
\end{equation}
and
\begin{equation}
\omega=\Omega_{i}\Biggl(1-\frac{8\pi
k_{z}^{2}c^{2}\chi_{i}}{\omega_{Li}^{2}-2k_{z}^{2}c^{2}+2\Omega_{i}^{2}}\Biggr),
\label{NBS new ion par branch}
\end{equation}
In this equation a follows notation is used $\omega_{La}^{2}=4\pi
e_{a}^{2}n_{0a}/m_{a}$ -- is the Langmuir frequency,
$\Omega_{c}=e_{c}B_{0}/(m_{c}c)$ that two parameters it is the cyclotron
frequency which arise from the motion of the charge $e_{c}$ and
the magnetic moment in the external magnetic field $B_{0}$
correspondingly, $\gamma_{a}$ is the gyromagnetic ratio, for
example, for neutrons $\gamma_{b}=-1.91\mu_{nuc}$, where
$\mu_{nuc}$ is the nuclear magneton.

The frequency dependence of this waves is located around the electron
or the ion cyclotron frequencies.

In the absence of the plasma we can obtain dispersion relations for
the eigenwaves in the neutron beam. There are three solutions:
\begin{equation}
\omega^{2}=k^{2}c^{2}\mp 4\pi\lambda k_{z}U_{z}W_{\gamma
b},\label{NBS light in NB}
\end{equation}

\begin{equation}
\omega=k_{z}U_{z}\pm\Omega_{\gamma b}-4\pi\lambda W_{\gamma b},
\label{NBS gen by SS,SP,SO}
\end{equation}

\begin{equation}
\omega=k_{z}U_{z}\mp\Omega_{\gamma
b}+2\pi\lambda(k_{z}U_{z}\mp\Omega_{\gamma b})\frac{\Omega_{\gamma
b}W_{\gamma b}}{k^{2}c^{2}}, \label{NBS gen by SO}
\end{equation}
here and below $\lambda$ is equal to 1 ($\lambda =1$) and indicate
the contribution of the spin-orbit interaction, $W_{\gamma
a}=\chi_{a}\Omega_{\gamma a}$, $\Omega_{\gamma
a}=2\gamma_{a}B_{0}/\hbar$, for charged particles we can use equality $\Omega_{\gamma
a}=\Omega_{a}$.

Formulas (\ref{NBS light in NB})  is a dispersion relation for the
light which propagates through the neutron beam. Solution (\ref{NBS gen
by SO}) arises in consequence of the spin-orbit interaction.

Now we consider the resonance interaction of mode of beam
(\ref{NBS gen by SS,SP,SO}), (\ref{NBS gen by SO}) with waves in the
plasma. We present results both for well-known waves and for the waves
obtained in this paper.

First of all, we illustrate this effect on example of  the fast magneto-sound
wave. Let us consider the generation of the fast magneto-sound
wave. The condition of the resonance interaction of the fast
magneto-sound wave and beam mode with dispersion presented by
formula (\ref{NBS gen by SS,SP,SO}) is
$\omega_{0}=k_{z}U_{z}+\Omega_{\gamma b}$. Under this condition
instabilities are arisen, increment of instabilities is presented
by formula
$$\delta\omega^{2}=-2\pi\omega_{0}\mid W_{\gamma
b}\mid\times$$
\begin{equation} \times\frac{\lambda\mid\Omega_{\gamma b}\mid-2\lambda
k_{z}U_{z}+2(k_{z}c)^{2}/\omega_{0}}{2\omega_{0}-\sum_{c}\omega_{Lc}^{2}\Omega_{c}/(\omega_{0}+\Omega_{c})^{2}}<0,
\label{NBS incr 1}
\end{equation}
where $\omega_{0}=\omega_{0}(k_{z})$ -- is the dispersion dependence of the
fast magneto-sound wave in the absence of beam. For the fast
magneto-sound wave where is $\omega_{0}(k)\subset(0,\mid\Omega_{e}\mid)$ and
$2\omega_{0}(k)-\sum_{c}\omega_{Lc}^{2}\Omega_{c}/(\omega_{0}(k)+\Omega_{c})^{2}>0$.

Resonance interaction of the wave (\ref{NBS new ion par branch})
with beam mode (\ref{NBS gen by SS,SP,SO}) on condition
$\Omega_{i}(1+\delta\Omega/\Omega_{i})=k_{z}U_{z}+\Omega_{\gamma
b}$ lead to the instability with
$$\delta\omega^{2}=-\pi\times\frac{1}{1+\frac{3}{8}\frac{\omega_{Le}^{2}}{\mid\Omega_{e}\mid\Omega_{i}}}\times\Biggl(2\chi_{i}(k_{z}c)^{2}$$
\begin{equation}
 +\mid W_{\gamma
b}\mid\biggl(\lambda\mid\Omega_{\gamma
b}\mid+2k_{z}c(k_{z}c/\Omega_{i}-\lambda U_{z}/c)\biggr)\Biggr)<0
\label{NBS incr 2}.
\end{equation}

Under condition $\omega_{A}(k)=k_{z}U_{z}-\Omega_{\gamma b}$
(where $\omega_{A}(k)$ the dispersion dependence of the Alfven wave in absence of the
beam, $\omega_{A}(k)\in(0,\Omega_{i})$) there is the generation of the Alfven waves
$$\delta\omega^{2}=-2\pi W_{\gamma b} \times$$
\begin{equation}
\times\frac{\lambda\omega_{A}(k) \Omega_{\gamma
b}+2(k_{z}c)^{2}-2\omega_{A}\lambda
k_{z}U_{z}}{2\omega_{A}(k)-\sum_{c}\omega_{Lc}^{2}\Omega_{c}/(\omega_{A}(k)+\Omega_{c})^{2}}<0.
\label{NBS Gen Alf Wave}
\end{equation}
at resonance with the beam mode (\ref{NBS gen by SS,SP,SO}).

In the case
$\mid\Omega_{e}\mid(1+\delta\Omega/\mid\Omega_{e}\mid)=k_{z}U_{z}+\Omega_{\gamma
b}$ there is the resonance interaction of the wave (\ref{NBS new el par
branch}) with the beam mode (\ref{NBS gen by SO}) leading to the frequency shift
\begin{equation}
\delta\omega=\pm\sqrt{\frac{32\pi\mid\Omega_{e}\mid
(\lambda\mid\Omega_{e}\Omega_{\gamma b}W_{\gamma
b}\mid-2k_{z}^{2}c^{2}\mid W_{\gamma
e}\mid)}{\omega_{Le}^{2}-8\Omega_{e}^{2}}}. \label{NBS 4th inr}
\end{equation}
For the dense plasma and conditions
$k_{z}^{2}c^{2}>\mid\chi_{b}/\chi_{e}\mid\Omega_{\gamma b}^{2}$,
$\mid\Omega_{\gamma e}\mid+\mid\Omega_{\gamma b}\mid=k_{z}U_{z}$
the solution (\ref{NBS 4th inr}) becomes imaginary, since condition for instabilities fulfills.

\section{ V. Propagation of waves perpendicular to magnetic field.}

In the case of the waves propagation perpendicular to the external magnetic field we consider only  the propagation of waves and do not consider the interaction with the beam. We show the
four new wave solutions arise in this case. Dispersion dependence of
obtained waves has form:
\begin{equation}\label{NBS Lambda zz tr new rel}\omega=|\Omega_{a}|\biggl(1-\frac{2\pi
k^{2}c^{2}\chi_{a}}{\omega^{2}_{e}+k^{2}c^{2}-\Omega^{2}_{a}}\biggr).\end{equation}

In this paper we also report about another two solutions. In
approximation of motionless ions one of presenting waves has follows
dispersion relation
$$\omega=\sqrt{\Omega_{e}^{2}+v_{qe}^{2}k_{\perp}^{2}}$$
\begin{equation}\label{NBS Lambda xy tr e only new rel 1}+8\pi^{2}\chi_{e}^{2}\frac{\Omega_{e}^{2}}{\sqrt{\Omega_{e}^{2}+v_{qe}^{2}k_{\perp}^{2}}}\times\frac{k_{\perp}^{4}c^{4}}{\omega_{e}^{2}(\omega_{e}^{2}+k_{\perp}^{2}c^{2})},\end{equation}
where
\begin{equation}\label{NBS quant vel} v_{qsa}^{2}=v_{sa}^{2}+\frac{\hbar^{2}k^{2}}{4m_{a}^{2}}.\end{equation}

Formulas (\ref{NBS Lambda zz tr new rel}), (\ref{NBS Lambda xy tr
e only new rel 1}) and all obtained in this paper solutions
exist only under the conditions $\chi_{a}\neq0$. They arise from
dispersion equations containing new term proportional  to
$\chi_{a}$.

\section{ VI. Spin waves}

Spin waves are the waves in which a perturbation of the electric
field has no influence on process of the wave propagation. This
waves propagate by means of the magnetic field.

Static magnetic field does not cause the electric field. When we consider a motion of charges or magnetic moments the magnetic field changes, it should lead to arises of the electric field according to the Maxwell's equations. At slow particles motion the magnetic field changes slowly. So, we can neglect the electric field which appears due to magnetic field changes. Charges also bring the electric field, but we suppose it give no contribution in the wave propagation. As a result we put the electric field in the set of equation equal to zero. If we put charge of particles equal to zero we get quasi-magnetostatic spin waves ~\cite{Maugin IJES 82}-~\cite{Erdem 07}. But in our paper we include a contribution of the charges.

Thus, for spin waves the amplitude of the electric field of wave
$\textbf{E}=0$, whereas the magnetic field in the wave is nonzero,
$\textbf{B}\neq0$. In system of equations (\ref{NBS momentum
balance eq}) and (\ref{NBS magn mom balance eq}), we assume
$\textbf{E}=0$, and obtain the following equation:
\begin{equation}\label{NBS disp matriks for sw}\Pi_{\alpha\beta}(\omega,\textbf{k})\delta B_{\beta}(\omega,\textbf{k})=0.\end{equation}
From equation (\ref{NBS disp matriks for sw}) we have dispersion equation
\begin{equation}\label{NBS disp eq for sw}\det\widehat{\Pi}=0.\end{equation}
We do not present explicit form of this equation because
of its largeness. Explicit form of the matrix $\Pi_{\alpha\beta}(\omega,\textbf{k})$
is presented in ~\cite{Andreev VestnMSU 2007}. Here we consider
influence of the spin-orbit interaction. This interaction realized
by means of electric field. Consequently, there is no contribution
of spin-orbit interaction in (\ref{NBS disp matriks for sw}). In
fact, we consider the equation obtained earlier ~\cite{Andreev VestnMSU 2007}.
But, we present more detailed analysis of this formula. Propagation of spin
waves along the direction of the external magnetic field was considered in
~\cite{Andreev VestnMSU 2007}. Here we study
propagation of the spin waves in the whole $\textbf{k}$ space.

\begin{figure}
\includegraphics[width=8cm,angle=0]{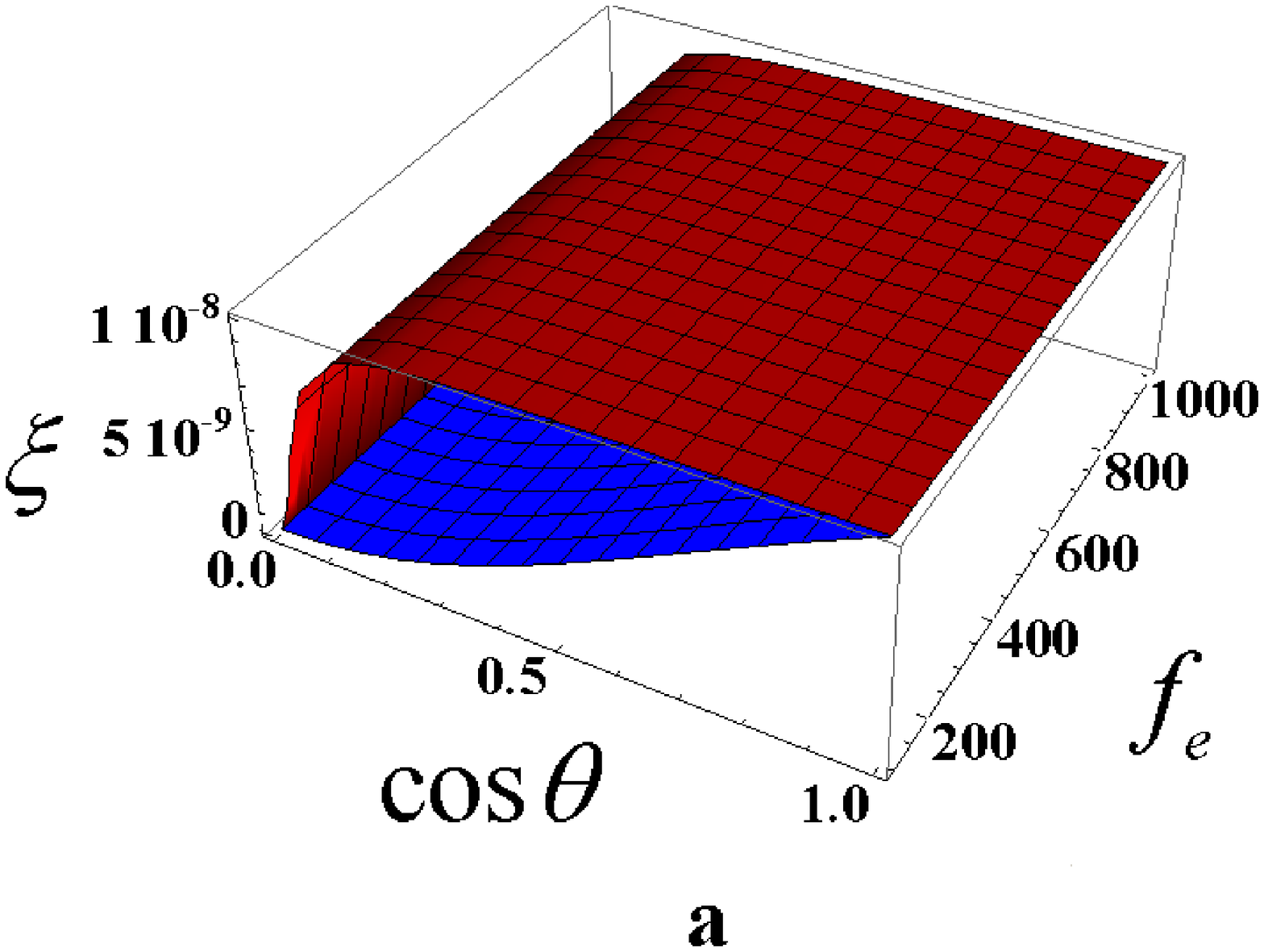}
\includegraphics[width=8cm,angle=0]{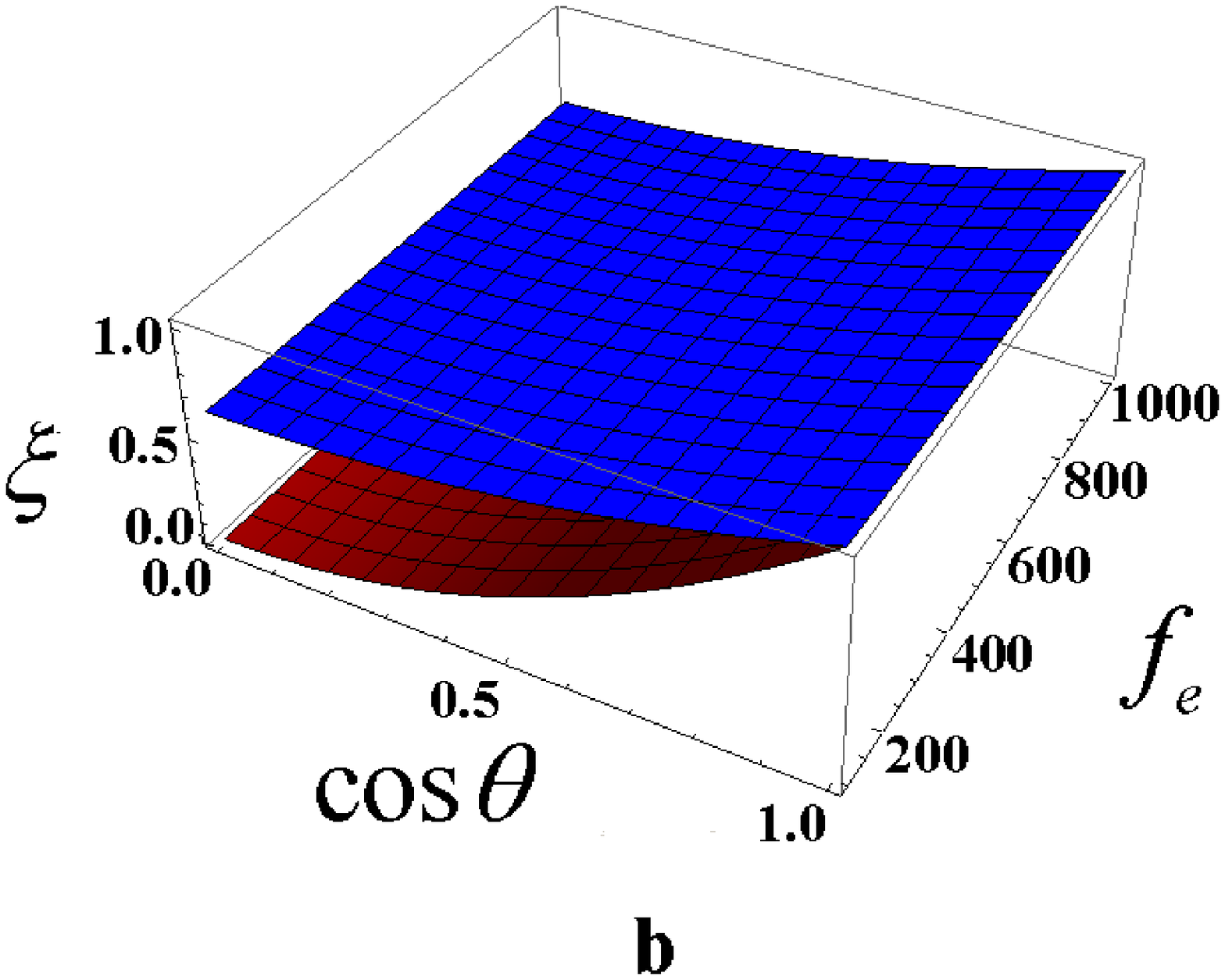}
\includegraphics[width=8cm,angle=0]{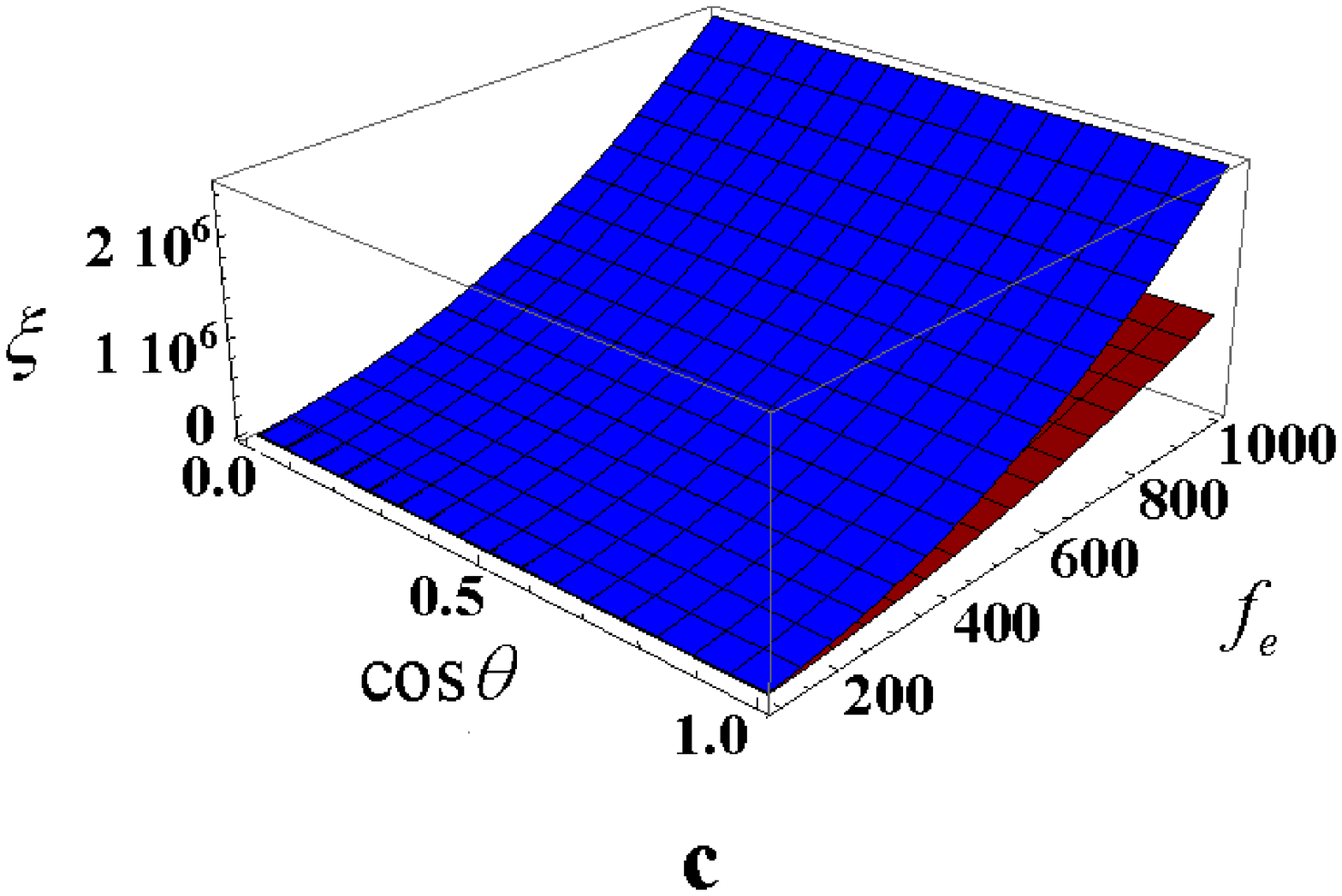}
\caption{\label{figSWS:epsart} This figure displays the dispersion
dependence determined by equation (\ref{NBS scsw hole k space lin
evident}) (in fact it is the function $\omega(k)$). Equation
(\ref{NBS scsw hole k space lin evident}) has three solutions. We
present this solution on fig a, b and c, correspondingly. Each
solution is obtained for two different values of
$\chi=\chi_{i}/\chi_{e}$. Here we interested in the cases when
$\chi_{e}$ and $\chi_{i}$ have the same order or differ by several orders. For room
temperature we suppose
$f_{i}\simeq 10^{-3}f_{e}$ and approximately take
$g\simeq 10^{-4}$. We consider the dense plasma. Therefore, the quantity $f_{e}$ is order of $10^{2}$-$10^{3}$. Thus, we consider $\xi$ like a
function of $k$ (via $f_{e}$) and $\cos\theta$:
$\xi=\xi(f_{e},\cos\theta)$. The parameter $\chi$ has a
following values. For the first case (blue surface) $\chi=1.5$
\textit{and} in the second case (red) $\chi=0.002$.}
\end{figure}

There are three solutions of dispersion equation $\det\hat{\Pi}=0$
for the case of the spin waves propagating parallel to the direction
of the external magnetic field. Two solutions are the oscillations with
constant frequencies:
\begin{equation}\label{NBS spin oscilation}\omega=|\Omega_{a}|(1-4\pi\chi_{a}).\end{equation}

At $m_{e}\chi_{i}\neq m_{i}\chi_{e}$ the solution is
\begin{equation}\label{NBS scsw new rel}\omega^{2}=k^{2}\frac{(v^{2}_{si}+\frac{\hbar^{2}}{4m^{2}_{i}}k^{2})\chi_{e}\mid\Omega_{e}\mid-(v^{2}_{se}+\frac{\hbar^{2}}{4m^{2}_{e}}k^{2})\chi_{i}\mid\Omega_{i}\mid}{\chi_{e}\mid\Omega_{e}\mid-\chi_{i}\mid\Omega_{i}\mid}.\end{equation}

This formula represents the dispersion of the self-consistent spin
waves in the system of the electrons and ions with nonzero intrinsic
magnetic moments.

Neglecting the quantum Bohm potential (the terms proportional
to $\hbar^{2}$), from (\ref{NBS scsw new rel}) we get
$$\omega=\tilde{v}_{sp}k,$$
where
$$\tilde{v}_{sp}=\frac{v^{2}_{si}\chi_{e}\mid\Omega_{e}\mid-v^{2}_{se}\chi_{i}\mid\Omega_{i}\mid}{\chi_{e}\mid\Omega_{e}\mid-\chi_{i}\mid\Omega_{i}\mid}$$
and $\tilde{v}_{sp}$ do not depend on wave vector $k$.

Let us consider the spin waves in the whole \textbf{k} space. Further we
write the evident form of the dispersion equation  (\ref{NBS disp eq for sw}).
We consider the terms proportional to
the magnetic susceptibility  $\chi_{a}$ in the first and the second degree. The dispersion equation is
\begin{widetext}
$$\sum_{a}\chi_{a}\frac{\Omega_{a}}{\Delta_{a}}\biggl(k_{z}^{2}(\omega^{2}-\Omega_{a}^{2})+k_{\perp}^{2}\omega^{2}\biggr)$$
\begin{equation}\label{NBS scsw hole k spese}+4\pi\sum_{a,b}\chi_{a}\chi_{b}\frac{1}{\Delta_{a}}\frac{\Omega_{a}\Omega_{b}}{\omega^{2}-\Omega_{b}^{2}}\biggl(2k_{z}^{2}\Omega_{b}(\omega^{2}-\Omega_{a}^{2})+k_{\perp}^{2}(\Omega_{a}+\Omega_{b})\omega^{2}\biggr)=0,\end{equation}
where
$$\Delta_{a}=\omega^{2}(\omega^{2}-\Omega^{2}_{a})-v^{2}_{qsa}(\omega^{2}k^{2}_{\perp}+(\omega^{2}-\Omega^{2}_{a})k^{2}_{z}).$$
\end{widetext}
Equation (\ref{NBS scsw hole k spese}) is the equation of the fifth degree of $\omega^{2}$.

The quantity $\chi_{a}$ is much less than 1: $\chi_{a}<<1$.
Consequently, in the simplest case, we neglect the terms
proportional to $\chi_{a}^{2}$, in comparison with the terms
proportional to $\chi_{a}$. We obtain equation of the third degree
of $\omega^{2}$.
\begin{equation}\label{NBS scsw hole k space lin}\sum_{a}\chi_{a}\frac{\Omega_{a}}{\Delta_{a}}(k_{z}^{2}(\omega^{2}-\Omega_{a}^{2})+k_{\perp}^{2}\omega^{2})=0\end{equation}
or in the explicit form
$$(\chi_{e}+\chi_{i}g)\xi^{3}$$
$$-\biggl((\chi_{e}+\chi_{i}g^{3})\cos^{2}\theta+\chi_{e}(g^{2}+f_{i}^{2})+\chi_{i}g(1+f_{e}^{2})\biggr)\xi^{2}$$
$$+\biggl(\chi_{e}(g^{2}+f_{i}^{2})\cos^{2}\theta+\chi_{i}g^{3}(1+f_{e}^{2})\cos^{2}\theta$$
\begin{equation}\label{NBS scsw hole k space lin evident}+g(\chi_{e}f_{i}^{2}g+\chi_{i}f_{e}^{2})\biggr)\xi -g^{2}\cos^{2}\theta(\chi_{e}f_{i}^{2}+\chi_{i}f_{e}^{2}g)=0,\end{equation}
where $\xi=\omega^{2}/\Omega_{e}^{2}$, $g=\Omega_{i}/\Omega_{e}$,
$f_{a}^{2}=v_{qsa}^{2}k^{2}/\Omega_{e}^{2}$, $\theta$ is the polar
angle: $\cos\theta=k_{z}/k$ and
$k=\sqrt{k_{z}^{2}+k_{\perp}^{2}}$. Equation (\ref{NBS scsw hole k
space lin evident}) is the third degree in $\xi$ and we can not
obtain analytical solutions of this equation. Consequently, we
present it's numerical solution on Fig. (1). We consider two type
of material, the first one with close values of $\chi_{e}$ and $\chi_{i}$ and the second one such that $\chi_{i}\ll\chi_{e}$.

For one kind of particle there is only one spin mode. It is
\begin{equation}\label{NBS cicl freq wh k sp}\omega=\mid\Omega_{a}\mid \cos\theta.\end{equation}
Solution (\ref{NBS cicl freq wh k sp}) is generalization of
(\ref{NBS spin oscilation}) for whole $\textbf{k}$ space.

The terms proportional to $\chi_{a}^{2}$ have significant
contribution in equation (\ref{NBS scsw hole k spese}) at
$\omega^{2}-\Omega_{a}^{2}\approx 0$. In this case we obtain
analytical solutions. Subindex $a$ equal $e$ and $i$. Since we
have two wave solutions, for each sorts of particles. For both waves the dispersion relation may be presented in the form:
$$\omega=\mid\Omega_{a}\mid+\delta\omega,$$
with $a=e,i$. For wave with frequency around the electron cyclotron
frequency we have
\begin{widetext}
\begin{equation}\label{NBS spin e WhKS}\delta\omega=\pi\chi_{e}\Omega_{e}\frac{2\chi_{e}\Omega_{e}(\Omega_{e}^{2}-v_{i}^{2}k^{2})-\chi_{i}\Omega_{i}v_{qSe}^{2}(2k_{z}^{2}+k_{\perp}^{2})}{\chi_{e}\Omega_{e}(\Omega_{e}^{2}-v_{i}^{2}k^{2})-\chi_{i}\Omega_{i}v_{qSe}^{2}k^{2}},\end{equation}
and around the ion cyclotron frequency
\begin{equation}\label{NBS spin i WhKS}\delta\omega=2\pi\chi_{i}\Omega_{i}\frac{-2\chi_{i}\Omega_{i}\biggl((\Omega_{e}^{2}k_{z}^{2}-\Omega_{i}^{2}k_{\perp}^{2})v_{qSe}^{2}-\Omega_{i}^{2}\Omega_{e}^{2}\biggr)+\chi_{e}\Omega_{e}^{2}(\Omega_{i}k_{\perp}^{2}-2\Omega_{e}k_{z}^{2})v_{i}^{2}}{\chi_{i}\Omega_{i}\biggl((\Omega_{e}^{2}k_{z}^{2}-\Omega_{i}^{2}k_{\perp}^{2})v_{qSe}^{2}-\Omega_{i}^{2}\Omega_{e}^{2}\biggr)+\chi_{e}\Omega_{e}(\Omega_{e}^{2}k_{z}^{2}-\Omega_{i}^{2}k_{\perp}^{2})v_{i}^{2}},\end{equation}
\end{widetext}
where $v_{qsa}^{2}$ defined by the formula (\ref{NBS quant vel}).

Under the condition $k_{\perp}=0$ solutions (\ref{NBS spin e WhKS})
and (\ref{NBS spin i WhKS}) pass to
$$\delta\omega=2\pi\chi_{e}\Omega_{e},$$
for (\ref{NBS spin e WhKS}), and
$$\delta\omega=-4\pi\chi_{i}\Omega_{i},$$
for (\ref{NBS spin i WhKS}).

The spin-orbit interaction has no influence on dynamic of the spin waves
in the plasma. We investigate the spin waves like waves in propagation
of which the electric field take no part. But force of the spin-orbit
interaction is proportional to the electric field. Problem of the
generation of the spin-waves which propagate parallel to the
direction of the external magnetic field was analyzed in ~\cite{Andreev AtPhys 08}.

In this paper we interested in effects existed at propagation of the
neutron beam through the plasma parallel to the external magnetic
field. In this case there are instabilities of the spin waves, but
waves propagated parallel to the external magnetic field, only. Three
waves exist at described conditions. Dispersion of this waves
presented by formulas (\ref{NBS spin oscilation}) and (\ref{NBS
scsw new rel}). For the last one there is no instability. For the
two solutions (\ref{NBS spin oscilation}) we obtain following
increments of instabilities.

In the absence of the medium, for the spin waves in the neutron beam we
get the dispersion relation.
\begin{equation}\label{NBS v f2}
\omega=kU_{z}\pm\Omega_{b}(1-4\pi\chi_{b}).
\end{equation}

Under the conditions of resonance
$kU_{z}+\varepsilon\Omega_{b}=|\Omega_{c}|(1-4\pi\chi_{c})$, here $\varepsilon=\pm 1$, the
dependence of frequency from wave vector is
\begin{equation}\label{NBS subst} \omega=|\Omega_{c}|(1-4\pi\chi_{c})+\delta\omega .\end{equation}

Using substitution (\ref{NBS subst}), we receive $\delta\omega$ from the dispersion
equation. First of them is
\begin{equation}\label{NBS r f1}
\delta\omega^{2}=-(4\pi)^{2}\chi_{b}\chi_{c}\mid\Omega_{b}\mid\Omega_{c}.\end{equation}
This formula is valid for the electrons at condition
$\varepsilon=-1$, and for the ions at $\varepsilon=+1$. Second
solution has form
\begin{equation}\label{NBS r f2}
\delta\omega^{2}=-4\frac{\chi_{b}}{\chi_{c}}\mid\Omega_{b}\mid\Omega_{c},
\end{equation}
and valid for the ions at $\varepsilon=-1$ and for the electrons at
$\varepsilon=+1$. From formulas (\ref{NBS r f1}), (\ref{NBS r f2})
we see the solution for the ions wave take place just in case
paramagnetic ions.

It is the same results we obtained in paper ~\cite{Andreev AtPhys
08}, because the spin-orbit interaction has no influence on the spin
waves and their instabilities. Instabilities of the spin waves there
are due to the spin-spin and the spin-current interactions. This
interaction holds between magnetic moments of the beam
\textit{and} electric currents and magnetic moments of the plasma.

\section{ VII. Conclusion}

In this work we consider the influence of the magnetic moment or spins
of particles on dynamics of the magnetized plasma.

We discovered two new waves propagating in the plasma along the external
magnetic field. We obtained the four new wave propagating in the plasma perpendicular to the
direction of the magnetic field. We studied possibility of existence of the spin waves and obtained
the dispersion of the spin waves.

We show existence of resonant interaction of the neutron beam with the
plasma. This effect leads to instabilities in the plasma.
Consequently, there is generation of the waves in the plasma by means of the
neutron beam.

For generation of waves arisen from dynamic of magnetic moments
the neutron beam is more useful in compare with the beam of electrons.
It is because the beam of electrons, first of all, excites the waves of charge.

\section{ Appendix}

In section II, we present the Schrodinger equation contained only the
Coulomb interaction. In this paper we interested in the spin-spin, the
spin-current and the spin-orbit interactions. The whole Hamiltonian, used
in this work reads
\begin{widetext}
$$\hat{H}=\sum_{p}\biggl(\frac{1}{2m_{p}}\tilde{D}^{\alpha}_{p}\tilde{D}^{\alpha}_{p}+e_{p}\varphi^{ext}_{p}-\gamma_{p}\sigma^{\alpha}_{p}B^{\alpha}_{p(ext)}\biggr)-\sum_{p}\frac{\gamma_{p}}{m_{p}c}\varepsilon^{\alpha\beta\gamma}\sigma_{p}^{\alpha}E_{p,ext}^{\beta}\hat{D}_{p}^{\gamma}
$$
\begin{equation}\label{NBS ham gen}+\frac{1}{2}\sum_{p,n\neq p}(e_{p}e_{n}G_{pn}-\gamma_{p}\gamma_{n}G^{\alpha\beta}_{pn}\sigma^{\alpha}_{p}\sigma^{\beta}_{n})+\sum_{p,n\neq p}\frac{e_{p}}{m_{n}c}\lambda\gamma_{n} \varepsilon^{\alpha\beta\gamma}\partial_{n}^{\gamma}G_{pn}\sigma_{n}^{\beta}\hat{D}_{n}^{\alpha},\end{equation}
where
$$D_{p}^{\alpha}=-\imath\hbar\partial_{p}^{\alpha}+\frac{e_{p}}{c}A_{p,ext}^{\alpha},$$
\begin{equation}\label{NBS proizv} (\tilde{D}^{\alpha}_{p}\psi)_{s}(R,t)=\Biggl(\biggl(\frac{\hbar}{\imath}\partial^{\alpha}_{p}-\frac{e_{p}}{c}A^{\alpha}_{p (ext)}-\frac{e_{p}}{c}\sum_{n\neq p}(\frac{\xi}{2})\varepsilon^{\alpha\beta\gamma}\frac{r^{\beta}_{np}}{r^{3}_{np}}\gamma_{n}\sigma^{\gamma}_{n}\biggr)\psi\Biggr)_{s}(R,t), \end{equation}
\end{widetext}
here $\xi$ is an arbitrary numerical parameter. We choose the value of $\xi$ so
that the superposition principle of fields holds. We consider
superposition of the external magnetic field \textit{and} the field caused
by the magnetic moments and the electric currents. From calculation we get
$\xi=2$.

Here, we describe the meaning of terms in the Hamiltonian (\ref{NBS
ham gen}). The first term has a complex structure. This is the
kinetic energy of the particles, where we include vector potential
of magnetic field caused by spins, along with the external field.
Thereby, the first term include the spin-current interaction. The
second and third terms are the action of the external field on the charges and the
magnetic moments. Next term is effect of the external electric field
on moving magnetic moments or the spin-orbit interaction with the external
electric field ~\cite{Landau 4}. The first and second terms in the
second line are the Coulomb and the spin-spin interaction. And, the last
one is the spin-orbit interaction between particles.

The Green's functions of the Coulomb, the spin-spin and the spin-current interactions has the following form
$G_{pn}=1/r_{pn}$,
 $G^{\alpha\beta}_{pn}=4\pi\delta^{\alpha\beta}\delta(\textbf{r}_{pn})+\partial^{\alpha}_{p}\partial^{\beta}_{p}(1/r_{pn})$,
$C^{\alpha\beta}_{pn}=(e_{n}/c)\varepsilon^{\alpha\beta\gamma}r^{\gamma}_{pn}/r^{3}_{pn}$,
where $\gamma_{p}$ - is the gyromagnetic ratio. For electrons
$\gamma_{p}$ reads $\gamma_{p}=e_{p}\hbar/(2m_{p}c)$,
$e_{p}=-|e|$. The quantities
$\varphi^{ext}_{p}=\varphi(\textbf{r}_{p},t),A^{\alpha}_{p
(ext)}=A^{\alpha}(\textbf{r}_{p},t)$ are the scalar and the vector
potentials of the external electromagnetic field:
 $$ B^{\alpha}_{
(ext)}(\textbf{r}_{p},t)=\varepsilon^{\alpha\beta\gamma}\nabla^{\beta}_{p}A^{\gamma}_{(ext)}(\textbf{r}_{p},t),
$$
$$E^{\alpha}_{(ext)}(\textbf{r}_{p},t)=-\nabla^{\alpha}_{p}\varphi_{ext}(\textbf{r}_{p},t)-
 \frac{1}{c}\frac{\partial}{\partial t}A^{\alpha}_{ext}(\textbf{r}_{p},t) .$$
$\sigma^{\alpha}_{p}$ is the Pauli matrix, a commutation relations
for them is
$$[\sigma^{\alpha}_{p},\sigma^{\beta}_{n}]=2\imath\delta_{pn}\varepsilon^{\alpha\beta\gamma}\sigma^{\gamma}_{p}.$$

The Hamiltonian (\ref{NBS ham gen}) is an analog of the Breit's
Hamiltonian ~\cite{Landau 4} and ~\cite{spin-spin interaction
small}. Here we do not consider the semi-relativistic contribution in the
kinetic energy $\sim \textbf{D}_{p}^{4}$ and the current-current
interaction (the Biot-Savart law). Non-quantum part of the last
one we include by means of superposition principle of fields.

Using the Hamiltonian (\ref{NBS ham gen}) we obtained the QHD equation
contained many-particle correlations. This correlations include
the exchange interaction. We make notice here about the exchange
interaction existing for all kind of forces between particles. They are the Coulomb, the spin-spin, the spin-current and the spin-orbit
interactions. The equations of the momentum balance and magnetic moment evolution arise in the form of integro-differential equations. This form useful for studying low
dimensional systems (see for example ~\cite{Andreev arxiv 11 3}).
In this work we neglected correlations and use the
self-consistent field approximation. We made also the additional
approximations. We account terms coincided to the following non-linear
one-particle Schrodinger equation
$$\imath\hbar\partial_{t}\Phi(\textbf{r},t)=\frac{1}{2m}\widehat{\textbf{D}}^{2}\Phi(\textbf{r},t)-\mu\widehat{\sigma}^{\alpha}B^{\alpha}(\textbf{r},t)\Phi(\textbf{r},t)$$
$$+e\varphi(\textbf{r},t)\Phi(\textbf{r},t)-\frac{\gamma}{mc}\varepsilon^{\alpha\beta\gamma}\widehat{\sigma}^{\alpha}E^{\beta}(\textbf{r},t)\widehat{D}^{\gamma}(\textbf{r},t)\Phi(\textbf{r},t),$$
where
$$\widehat{\textbf{D}}=\widehat{\textbf{D}}(\textbf{r},t)=\widehat{\textbf{p}}-(e/c)\textbf{A}(\textbf{r},t)$$
and
$$\Phi^{*}(\textbf{r},t)\Phi(\textbf{r},t)=n(\textbf{r},t).$$

Here we would like to justify the choice of form of the Hamiltonian (\ref{NBS ham gen}),
especially of the first term contained the spin-current interaction.
For this purpose we consider classic Lagrangian. For one charged
particle particle in the external electromagnetic field the Lagrangian has form:

$$L=\frac{m\textbf{v}^{2}}{2}-e\varphi+\frac{e}{c}\textbf{v}\textbf{A}.$$
For particles having magnetic moment there is additional term to the potential energy, this term has form
$$\Delta U=-\mu^{\alpha}B^{\alpha}.$$
In this case the  Lagrangian is
$$L=\frac{m\textbf{v}^{2}}{2}-e\varphi+\frac{e}{c}\textbf{A}\textbf{v}+\mu^{\alpha}B^{\alpha}.$$
Starting from this point we can build Lagrangian for $N$ interacting particles
$$L=\sum_{i}\frac{m_{i}\textbf{v}_{i}^{2}}{2}-\sum_{i,j\neq i}e_{i}e_{j}G_{ij}$$
\begin{equation}\label{NBS Lagr many part}+\sum_{i}\biggl(\frac{e_{i}}{c}\textbf{v}_{i}\textbf{A}_{i}+\mu^{\alpha}_{i}B^{\alpha}_{i}\biggr),\end{equation}
where $\textbf{A}_{i}$, $\textbf{B}_{i}$ describe magnetic fields
made by $j$-th particles and act on $i$-th particle.  We do not
consider the external field for compactness. The sources of the magnetic
field are the moving charges
$$\textbf{A}_{i}=\sum_{j\neq i}\frac{e_{j}}{c}\frac{\textbf{v}_{j}}{r_{ij}},$$
and the magnetic moments
$$A^{\alpha}_{i}=-\sum_{j\neq i}\varepsilon^{\alpha\beta\gamma}\frac{r_{ij}^{\alpha}}{r_{ij}^{3}}\mu_{j}^{\gamma}.$$
We do non considered the current-current interaction, which is proportional to the square
of the velocity. Magnetic field has form
$$B^{\alpha}_{i}=\sum_{j\neq i}G_{ij}^{\alpha\beta}\mu_{j}^{\beta}$$
caused by magnetic moments and
$$B^{\alpha}_{i}=-\sum_{j\neq i}\varepsilon^{\alpha\beta\gamma}\frac{e_{j}}{c}\frac{r_{ij}^{\beta}}{r_{ij}^{3}}v_{j}^{\gamma}$$
caused by moving chargees. The quantity $G_{ij}^{\alpha\beta}$ is
the Green's function of the spin-orbit interaction presented above. In
this way the Lagrangian (\ref{NBS Lagr many part}) is

$$L=\sum_{i}\biggl(\frac{m_{i}\textbf{v}_{i}^{2}}{2}-e_{i}\varphi_{i}^{ext}+\frac{e_{i}}{c}\textbf{v}_{i}\textbf{A}_{i}^{ext}+\mu^{\alpha}_{i}B_{i}^{\alpha ext}\biggr)$$

$$+\sum_{i,j\neq i}\biggl(-\frac{1}{2}e_{i}e_{j}G_{ij}+\frac{1}{2}G_{ij}^{\alpha\beta}\mu_{i}^{\alpha}\mu_{j}^{\beta}+\frac{e_{i}}{c}\varepsilon^{\alpha\beta\gamma}\mu_{j}^{\alpha}r_{ij}^{\beta}v_{i}^{\gamma}\frac{1}{r_{ij}^{3}}\biggr),$$
here we include external field. The corresponding Hamiltonian is
$$H=\sum_{i}\frac{1}{2m_{i}}\biggl(\textbf{p}_{i}-\frac{e_{i}}{c}\textbf{A}_{i}^{ext}-\frac{e_{i}}{c}\sum_{j\neq i}\varepsilon^{\alpha\beta\gamma}\mu^{\beta}_{j} r^{\gamma}_{ij}\frac{1}{r_{ij}^{3}}\biggr)^{2}$$

\begin{equation}\label{NBS ham classic}+\sum_{i}(e_{i}\varphi_{i}^{ext}-\mu^{\alpha}_{i}B_{i}^{\alpha ext})+\frac{1}{2}\sum_{i,j\neq i}\biggl(e_{i}e_{j}G_{ij}-\mu_{i}^{\alpha}\mu_{j}^{\beta}G_{ij}^{\alpha\beta}\biggr).\end{equation}
Using the usual quantization method for the Hamiltonian (\ref{NBS ham
classic}) we obtain (\ref{NBS ham gen}), except the spin-orbit
interaction.


\begin{thebibliography}{99}


\bibitem{Miller BOOK} R. B. Miller, \textit{An Introduction to the Physics of Intense Charged
Particle Beams} (Plenum, New York, 1982).

\bibitem{Ichimaru BOOK} S. Ichimaru, \textit{Basic Principles of Plasma Physics} (W. A. Benjamin,
Inc., Reading, Massachusetts, 1973).

\bibitem{Mikhailovskii BOOK}A. B. Mikhailovskii,
\textit{Theory of Plasma Instabilities} (Consultant
Bureau, New York, 1974).

\bibitem{Bret PRL 08} A. Bret, L. Gremillet, D. Benisti, and E. Lefebvre,
Phys. Rev. Lett. \textbf{100}, 205008 (2008).

\bibitem{Bret PRE 04} A. Bret, M.-C. Firpo, and C. Deutsch,
Phys. Rev. E \textbf{70}, 046401 (2004).

\bibitem{Gremillet PP 07} L. Gremillet, D. Benisti, E. Lefebvre, A. Bret,
Physics of plasmas \textbf{14}, 040704 (2007).

\bibitem{Nakar AstroP 11} E. Nakar, A. Bret, and M. Milosavljevic,
The Astrophysical Journal, \textbf{738}, 93 (2011).

\bibitem{Bret PP 06} A. Bret, C. Deutsch, Physics of plasmas \textbf{13}, 042106 (2006).

\bibitem{Timofeev PP 09} I. V. Timofeev, K. V. Lotov, and A. V. Terekhov, Physics
of plasmas \textbf{16}, 063101 (2009).

\bibitem{Bret PP 11} A. Bret and F. Haas, Physics of plasmas \textbf{18}, 072108 (2011).

\bibitem{Bret PP 08} A. Bret and M. E. Dieckmann, Physics of plasmas \textbf{15}, 062102 (2008).

\bibitem{P.K. Shukla UFN 10} P. K. Shukla, B. Eliasson, Phys. Usp. \textbf{53} 51
(2010) [Uspehi Fizihceskih Nauk \textbf{180}, 55 (2010)].

\bibitem{Shukla RMP 11} P. K. Shukla, B. Eliasson,
Rev. Mod. Phys. \textbf{83},  885 (2011).

\bibitem{Andreev arxiv 11 3} P. A. Andreev, L. S. Kuz'menkov, M. I. Trukhanova, Phys. Rev. B \textbf{84}, 245401 (2008).

\bibitem{Asenjo arxiv 2011} F. A. Asenjo, J. Zamanian, M. Marklund, G. Brodin,
and P. Johansson, arXiv:1108.4781.

\bibitem{Maksimov TMP 2002} L. S. Kuz'menkov and S. G. Maksimov,  Teor. i Mat. Fiz.,
\textbf{131} 231 (2002) [Theoretical and Mathematical Physics \textbf{131} 641 (2002)].

\bibitem{Tyshetskiy 2011} Yu. Tyshetskiy, S.V. Vladimirov, R. Kompaneets, arXiv:1108.0988.

\bibitem{MaksimovTMP 1999} L. S. Kuz'menkov and S. G. Maksimov,  Teor. i Mat. Fiz.,
 \textbf{118} 287 (1999) [Theoretical and Mathematical Physics \textbf{118} 227 (1999)].

\bibitem{MaksimovTMP 2001} L. S. Kuz'menkov, S. G. Maksimov, and V. V. Fedoseev, Theor.
Math. Fiz. \textbf{126} 136 (2001) [Theoretical and Mathematical
Physics, \textbf{126} 110 (2001)].

\bibitem{Marklund PRL07} M. Marklund and G. Brodin,
Phys. Rev. Lett. \textbf{98}, 025001 (2007).

\bibitem{Asenjo PP 11} F. A. Asenjo, V. Muáoz, J. A. Valdivia, and S. M.
Mahajan, Phys. Plasmas \textbf{18}, 012107 (2011).

\bibitem{Andreev PRA08} P. A. Andreev, L. S. Kuz'menkov, Phys. Rev. A \textbf{78}, 053624 (2008).

\bibitem{Andreev AtPhys 08} P. A. Andreev, L.S.  Kuz'menkov,
Physics of Atomic Nuclei \textbf{71},
N.10, 1724 (2008).

\bibitem{Andreev VestnMSU 2007} P. A. Andreev, L.S. Kuz'menkov,
Moscow
University Physics Bulletin \textbf{62}, N.5, 271 (2007).

\bibitem{Andreev PIERS 2011} P. A.  Andreev  and L. S. Kuz'menkov,
 PIERS Proceedings, pp.1047, March 20-23, Marrakesh,
MOROCCO 2011.

\bibitem{Maugin IJES 82} G. A. Maugin, A. Fomethe, International Journal of Engineering Science
\textbf{20}, 885 (1982).

\bibitem{Barybin EPJAP 03} A. A. Barybin, The European Physical Journal Applied Physics
\textbf{22}, 189 (2003).

\bibitem{Erdem 07} A. Unal Erdem, Vibration Problems ICOVP 2005,
Springer Proceedings in Physics  \textbf{111}, 175 (2007).

\bibitem{Landau 4} V.B. Berestetskii, E.M. Lifshitz, L.P. Pitaevskii, \emph{Quantum
Electrodynamics}, Vol. 4, 2nd ed. (Butterworth-Heinemann, 1982).

\bibitem{spin-spin interaction small}  G. Breit, Phys. Rev. \textbf{34}, 553
(1929);  V. Yu. Lazur, S. I. Myhalyna, and O. K. Reity, Phys. Rev.
A \textbf{81}, 062707 (2010).




\end{thebibliography}
\end{document}